\documentclass[aps,prd,showkeys,showpacs,twocolumn]{revtex4}

\usepackage[colorlinks=true,
            linkcolor=blue,
            urlcolor=blue,
            citecolor=blue]{hyperref}
            
\usepackage{amssymb,amsmath,epsfig}
\usepackage{eurosym}
\usepackage{amsfonts}
\usepackage{array}
\usepackage{changes}
\usepackage{amsthm}
\usepackage{bm}
\usepackage{palatino}
\def\0{{\sst{(0)}}}
\def\1{{\sst{(1)}}}
\def\2{{\sst{(2)}}}
\def\3{{\sst{(3)}}}
\def\4{{\sst{(4)}}}
\def\5{{\sst{(5)}}}
\def\6{{\sst{(6)}}}
\def\7{{\sst{(7)}}}
\def\8{{\sst{(8)}}}
\def\sst#1{{\scriptscriptstyle #1}}

\usepackage{mathpazo}
\usepackage{supertabular}
\usepackage{graphics}
\usepackage{color}
\usepackage{graphicx}

\begin{document}

\title{Black Hole Shadow in Symmergent Gravity}

\author{{\.I}rfan {\c C}imdiker$^a$, Durmu{\c s} Demir$^b$ and Ali {\"O}vg{\"u}n$^c$} \email{ilimirfan@ktu.edu.tr}

\affiliation{${}^a$Department of Physics, Karadeniz Technical University, 61080 Trabzon, Turkey}
\affiliation{${}^b$Faculty of Engineering and Natural Sciences, Sabancı University, 34956 İstanbul, Turkey}
\affiliation{${}^c$Physics Department, Eastern Mediterranean University, Famagusta, 99628 North Cyprus, via Mersin 10, Turkey}

\begin{abstract}
Symmergent gravity is the $R+R^2$ gravity theory which emerges in a way restoring gauge symmetries broken explicitly by the ultraviolet cutoff in effective field theories. To test symmergent gravity  we construct novel black hole solutions in four dimensions, and  study their shadow in the vacuum as well as plasma medium. Our detailed analyses show that the horizon radius, Hawking temperature, Bekenstein-Hawking entropy, shadow angular radius, and photon deflection angle are sensitive probes of the symmergent gravity and particle spectrum of the underlying quantum field theory. %Our exact black hole solution in symmergent gravity can be tested in Event Horizon Telescope (EHT) and other observations such as gravitational waves from binary black hole mergers observed by the Laser Interferometer Gravitational-wave Observatory (LIGO).  
\end{abstract}
\date{\today}


\maketitle

\section{Introduction}
The question of how quantum fields gravitate is a long-standing fundamental problem. If gravity is not quantum, which may indeed be the case \cite{dyson,thooft}, its incompatibility \cite{incompatible,wald} with quantum fields  might be an indication of its induced or emergent nature \cite{sakharov,visser,verlinde}. Recently, Sakharov's induced gravity \cite{sakharov} has been extended to gauge sector,  and gravity has been found to  emerge in way erasing anomalous gauge boson masses \cite{demir1} and thereby naturalizing the effective field theories \cite{weinberg,eff-action2}. 
The resulting setup \cite{demir1,demir2},  the Einstein-Hilbert term plus quadratic curvature terms plus dimensionally-regularized quantum field theory (QFT), possesses novel features that can be probed via collider, astrophysical and cosmological phenomena. This emergent gravity, termed as symmergent gravity for its gauge  symmetry-restoring nature \cite{demir2,demir3}, has all its couplings induced by flat spacetime quantum loops, and therefore, its underlying QFT can be probed or constrained via various phenomena.  It should be emphasized that symmergent gravity is not an effective field theory constructed in curved spacetime \cite{visser,birrel}. It is flat spacetime effective field theory taken to curved spacetime in a way restoring gauge symmetries broken explicitly by the UV cutoff overlying the QFT \cite{demir1,demir2}. As a matter of fact, quadratic curvature terms in symmergent gravity, different than those in the other approaches \cite{visser,birrel}, have recently been shown to lead to successful Starobinsky inflation \cite{irfan}.

In view of its widespread effects, in the present work we study formation and shadow angular radius of spherically symmetric black holes (BHs) in symmergent gravity. The symmergent gravity falls in the $f(R)$ gravity  \cite{f(R)review1,f(R)review2,f(R)review3} classification, where analyses have already been carried out of the formation and signatures of BHs \cite{Ovgun:2018gwt,Ovgun:2017bgx,Banerjee:2017nib,Nashed:2020mnp,Halilsoy:2015qta}, with further work \cite{Ovgun:2018tua,Ovgun:2020gjz,Javed:2020mjb,shadow} on their shadows. Imaging of the M87$^\star$ supermassive BH by the Event Horizon Telescope (EHT) \cite{eht} has started a new era in which it will be possible to test strong gravity regime and also probe the near horizon geometry of BHs \cite{eht,Herdeiro:2021lwl,Cunha:2020azh,Cunha:2019dwb,Cunha:2018gql,Cunha:2018acu,Falcke:1999pj,Younsi:2016azx,Hioki:2009na,Grenzebach:2014fha,Amarilla:2011fx,Abdujabbarov:2016hnw,Atamurotov:2013sca,Bambi:2013nla,Johannsen:2015mdd,Bambi:2019tjh,Vagnozzi:2019apd,Yin:2021fsg,Abbas:2021whh,Bouhmadi-Lopez:2019hpp,Rincon:2021hjj,Ghosh:2020cob,Shaikh:2018lcc,Shaikh:2019fpu,Konoplya:2019sns,Wang:2017hjl,Konoplya:2019fpy,Contreras:2019cmf,Contreras:2019nih,Akiyama:2019cqa,Gralla:2020pra,Johannsen:2015hib,Abbott:2020jks,Belhaj:2020rdb,Zhang:2019glo,Hendi:2020ebh,Konoplya:2020bxa,synge,Luminet:1979nyg,Brahma:2020eos}. Moreover, the next mission for the EHT, scheduled to start in near future, will release the image of our galactic SgrA* supermassive BH. 

In the present work we report on new exact BH solutions in symmergent gravity, with detailed investigation of  its shadow angular radius and thermodynamics. We determine  constraints on the QFT underlying the symmergent gravity using the entropy, temperature and shadow cast. We conclude that symmergent gravity could be 
tested observationally in near future in, for instance, ultra-high energy cosmic rays, black holes shadows, deflection angle, and other phenomena pertaining to strong gravity. 

The layout of the paper is as follows: First, we briefly review the symmergent gravity. Second, we construct spherically symmetric BH solutions in symmergent gravity, and determine constraints on the underlying QFT by using the BH shadow angular radius, weak deflection angle, entropy and temperature. Finally, we conclude.

\section{Symmergent Gravity}
The standard model of elementary particles (SM),  experimentally completed by the discovery of Higgs boson at the LHC \cite{higgs}, needs be  extended for various reasons such as dark matter, gravity, neutrino masses, and the like. In need of a general extension, it proves convenient to consider a generic renormalizable QFT made up of the SM particles and new particles beyond it. The   SM and the particles beyond it do not have to interact. In general, in accordance with Weinberg's effective field theory approach \cite{weinberg}, the QFT holds good up to some  physical UV momentum cutoff $\Lambda$. The fundamental difficulty is that this effective QFT exhibits a destructive sensitivity to the UV physics \cite{weinberg2} in scalar masses, gauge boson masses  as well as the vacuum energy. In explicit terms, the effective QFT  is endowed a power-law correction part \cite{weinberg2,eff-action2,ccb1}
\begin{widetext}
\begin{eqnarray}
S_{power}= \int d^4 x\ \sqrt{-\eta}\left\{-c_S \Lambda^2 S^2 - {c_V} \Lambda^2 {\mbox{tr}}\!\left[\eta_{\mu\nu} V^{\mu} V^{\nu}\right] - \sum_m c_m \Lambda^2 m^2 - c_O \Lambda^4\right\}
\label{power-law}
\end{eqnarray}
\end{widetext}
in which $\eta_{\mu\nu}={\rm diag.}(1,-1,-1,-1)_{\mu\nu}$ is the flat  metric, and $c_O$,
$c_m$, $c_S$ and $c_V$ are loop factors (coefficients of the powers of $\Lambda$).  The sum $\sum_m$ runs over all the particle masses in the QFT, with $\sum_m c_m m^2 \propto {\rm str}[m^2]$ at one loop. In general, $c_O \propto n_B-n_F$ where $n_B$ ($n_F$) is the total number of bosons (fermions) in the QFT \cite{demir1,demir2,demir3}. The power-law corrections in (\ref{power-law}) render the QFT unphysical from various aspects. First, the scalars $S$ acquire $\Lambda$-sized masses, which become  unnaturally large at large $\Lambda$ (the big hierarchy problem  \cite{ghp1,ghp2,ghp3}).  Second, the gauge bosons $V_{\mu}$ develop  $\Lambda$-sized anomalous masses, which  break all gauge symmetries explicitly (the charge and color breaking problem \cite{ccb1,ccb2}). Third,  the vacuum gains a $\Lambda^4$-sized energy density, which would cause the cosmological constant problem  \cite{ccp1,ccp2} if the spacetime were curved. It is  after the solution of these destructive problems that the effective QFT can gain physical relevance \cite{weinberg}.  It is clear that solutions based on not-weakly-coupled new particles (like superpartners in supersymmetry) may not be realistic because ATLAS and CMS experiments have detected so far no new particles \cite{exotica}. 

\subsection{Gauge Symmetry Restoration by Curvature}
The effective action (\ref{power-law}) gives the gauge boson anomalous mass action 
\begin{eqnarray}
S_V\left(\eta,\Lambda^2\right) &= -\int d^4x \sqrt{-\eta} {c_V} \Lambda^2 {\mbox{tr}}\!\left[\eta_{\mu\nu} V^{\mu} V^{\nu}\right] \label{deltSV}
\end{eqnarray}
in which ${\mbox{tr}}[\dots]$ stands for trace over group space. 

The gauge boson mass action (\ref{deltSV}) breaks gauge symmetries explicitly. It must be deactivated  for massless gauge particles to remain massless and massive gauge particles to receive their masses from spontaneous symmetry breaking. The usual way of restoring gauge symmetries is to introduce spurions \cite{spurion1} -- non-dynamical fields which realize the broken symmetries \cite{spurion1,spurion2,spurion3}. This means that one takes $\Lambda^2$ in (\ref{deltSV}) to a scalar field $L^2(x)$
\begin{eqnarray}
\Lambda^2 \longrightarrow L^2(x)
\label{spur-map}
\end{eqnarray}
such that $L^2(x)$ will eventually lead to the restoration of the gauge symmetries. To see how this said restoration takes place  it proves useful to start with the trivial identity 
\begin{eqnarray}
S_V\left(\eta, \Lambda^2\right)= S_V\left(\eta, \Lambda^2\right) - I_V(\eta) + {\tilde{I}}_V(\eta) 
\label{1st}
\end{eqnarray}
based on the two gauge-invariant kinetic constructs 
\begin{eqnarray}
I_V &=& \int d^{4}x \sqrt{-\eta}
\frac{c_V}{2} {\mbox{tr}}\!\left[V^{\mu\nu} V_{\mu\nu}\right]\nonumber\\ {\tilde{I}}_V &=& \int d^{4}x  \sqrt{-\eta} c_V {\mbox{tr}}\Big[V^{\mu}\!\left(
\!-D^2 \eta_{\mu\nu} + D_{\mu}D_{\nu}+iV_{\mu\nu}\!\right)\!V^{\nu}\nonumber\\ &+& {\partial}_{\mu} \left(\eta_{\alpha\beta} V^{\alpha} V^{\beta\mu}\right)\!\Big]
\label{IV-second}
\end{eqnarray}
which are identical ($I_V \equiv {\tilde{I}}_V$) under by-parts integration. Indeed, since ${\tilde{I}}_V$ aptly maintains both the surface and bulk terms $I_V$ and ${\tilde{I}}_V$ are always identical, with the gauge-covariant derivative $D_{\mu}$ and field strength tensor $V_{\mu\nu}$.

Now, using (\ref{IV-second}) for ${\tilde{I}}_V$ and (\ref{deltSV}) for $S_V$ at the right-hand side of (\ref{1st}) we get the identity
\begin{widetext}
\begin{eqnarray}
S_{V}\!\left(\eta, \Lambda_s^2\right) = 
 - I_V(\eta)  + \int d^{4}x  \sqrt{-\eta} c_V {\mbox{tr}}\left[V^{\mu}\!\left(
-D^2 \eta_{\mu\nu} - L^2 \eta_{\mu\nu} + D_{\mu}D_{\nu}+iV_{\mu\nu}\!\right)V^{\nu} + {\partial}_{\mu} \left(\eta_{\alpha\beta} V^{\alpha} V^{\beta\mu}\right)\right]
\label{2nd}
\end{eqnarray}
\end{widetext}
which is equal to (\ref{deltSV}) with $\Lambda^2$ replaced by $L^2(x)$. This equality is ensured by the identity
 $-I_V(\eta)+{\tilde{I}}_V(\eta) \equiv 0$. Bu this identity is specific to flat spacetime. It does not hold in curved spacetime. To see why, first one gets to curved spacetime of metric $g_{\mu\nu}$ 
using the general covariance map  \cite{covar}
\begin{eqnarray}
\eta_{\mu\nu}\rightarrow g_{\mu\nu}\,,\; \partial_\mu \rightarrow \nabla_\mu
\label{covar-map}
\end{eqnarray}
in which  $\nabla_\mu$ is the covariant derivative with respect to the Levi-Civita connection 
\begin{eqnarray}
{}^g\Gamma^\lambda_{\mu\nu}= \frac{1}{2} g^{\lambda\rho}\left( \partial_\mu g_{\nu\rho} + \partial_{\nu} g_{\rho\mu} - \partial_\rho g_{\mu\nu} \right) 
\label{LC}
\end{eqnarray}
Next, one discovers that 
\begin{eqnarray}
-I_V(g)+{\tilde{I}}_V(g) =-\int d^{4}x  \sqrt{-g} c_V {\mbox{tr}}\left[V^{\mu}R_{\mu\nu}V^\nu\right] 
\label{curved-id}
\end{eqnarray}
where $R_{\mu\nu}=R_{\mu\nu}({}^g\Gamma)$ is the Ricci tensor of the Levi-Civita connection. This result follows from the fact that ${\tilde{I}}_V(\eta)$, when taken to curved spacetime via the map (\ref{covar-map}), turns to ${\tilde{I}}_V(g)$ and ${\tilde{I}}_V(g)$ differs from $I_V(g)$ by the absence of 
the  $R_{\mu\nu}$ term (which would arise only in curved spacetime from the $[\nabla_\mu, \nabla_\nu]$ commutator). Then, under the identity (\ref{curved-id}) the flat spacetime spurion-enhanced gauge boson mass action in (\ref{2nd}) takes the ``curved" form 
\begin{widetext}
\begin{eqnarray}
S_{V}\left(g, \Lambda_s^2\right) =
 \int  d^{4}x  \sqrt{-g} c_V {\mbox{tr}}\left[V^{\mu}\left(
-L^2 g_{\mu\nu} - R_{\mu\nu}\right)V^{\nu}\right]
\label{3rd}
\end{eqnarray}
\end{widetext}
which is seen to vanish exactly if $L^2 g_{\mu\nu}$ equals the Ricci curvature $R_{\mu\nu}$. This, however, is simply impossible. The reason is that $R_{\mu\nu}$ does identically vanish in the flat limit ($g_{\mu\nu}\rightarrow \eta_{\mu\nu}$) but $L^2$ does not \cite{demir1,demir2,demir3}. The resolution is to make $L^2 g_{\mu\nu}$ dynamically approach to $R_{\mu\nu}$ namely as a solution of the equation of motion for $L^2$ (which is contributed  by all the terms in the power-law correction action in (\ref{power-law})) \cite{spurion3}. But $L^2$, as a spurion,  is devoid of any kinetic term and cannot therefore approach to $R_{\mu\nu}$ by its own dynamics. The resolution is to take $L^2$ itself a kinetic structure -- a scalar involving derivatives of a field. It turns out that the field in question can  be taken to be the affine connection $\Gamma^\lambda_{\mu\nu}$ \cite{spurion3}-- a general connection which has nothing to do with the Levi-Civita connection ${}^g\Gamma^\lambda_{\mu\nu}$ in (\ref{LC}) \cite{affine0,affine1}. The affine Ricci curvature 
\begin{eqnarray}
{\mathbb{R}}_{\mu\nu}(\Gamma) = \partial_\lambda \Gamma^\lambda_{\mu\nu} -\partial_\nu \Gamma^\lambda_{\mu\lambda} + \Gamma^\rho_{\lambda\rho} \Gamma^\lambda_{\mu\nu}  - \Gamma^\lambda_{\rho\nu}\Gamma^\rho_{\mu\lambda}
\label{a-Ricci}
\end{eqnarray}
involves derivatives of the affine connection \cite{affine0,affine1,affine2}, and $L^2$ itself becomes a kinetic structure by the identification (sign is opposite to that in \cite{demir1,demir2} due to opposite metric signature)
\begin{eqnarray}
\Lambda^2 g_{\mu\nu} = -{\mathbb{R}}_{\mu\nu}(\Gamma)
\label{spur-curv}
\end{eqnarray}
with which the gauge boson mass action (\ref{3rd}) takes the form
\begin{eqnarray}
\!S_{V}\!\left(g, {\mathbb{R}}\right)\! = \!\!
 \int\!\! d^{4}x  \sqrt{-g} c_V {\mbox{tr}}\!\left[V^{\mu}\!\left(\!
{\mathbb{R}}_{\mu\nu} - R_{\mu\nu}\!\right)\!V^{\nu}\right]
\label{3rdp}
\end{eqnarray}
after utilizing the equality (\ref{curved-id}). This metric-affine action \cite{affine2,affine3} gets deactivated or neutralized dynamically if the affine Ricci tensor ${\mathbb{R}}_{\mu\nu}\left(\Gamma\right)$ falls in close vicinity of the metrical Ricci tensor $R_{\mu\nu}({}^g\Gamma)$. As will be shown in subsection B below, $S_{V}\!\left(g, {\mathbb{R}}\right)$ does indeed vanish up to doubly Planck-suppressed terms and, in consequence, gauge symmetries get restored \cite{demir1,demir2,demir3}. 

Identification of $L^2$ with affine curvature in (\ref{spur-curv}) does actually imply an equivalence relation between $\Lambda^2$ and affine curvature. That relation comes out naturally via the spurion analysis and gauge symmetry restoration. But it has actually a symmetry reason. Indeed,  $\Lambda^2$ and curvature have a certain affinity in that $\Lambda^2$ breaks Poincare symmetry in flat spacetime \cite{Poincare-break,ccb1} and curvature does it in curved spacetime. It thus is not unexpected that they obey a certain affinity or equivalence relation \cite{demir1,demir2,demir3}. 

\subsection{Emergent General Relativity}
Under the successive maps (\ref{spur-map}) and (\ref{spur-curv}) the power-law corrections in (\ref{power-law}) lead to the following loop-induced curvature sector 
\begin{widetext}
\begin{eqnarray}
S_{curv}(g,{\mathbb{R}}) = \int d^4x \sqrt{-g}\left\{Q^{\mu\nu} {\mathbb{R}}_{\mu\nu}(\Gamma) + \frac{c_O}{16} \left(g^{\mu\nu} {\mathbb{R}}_{\mu\nu}(\Gamma)\right)^2 +  c_{V} {\mbox{tr}}\!\left[V_{\mu}V_{\nu}\right] R_{\mu\nu}({}^g\Gamma)\right\}
\label{action-affine-2px}
\end{eqnarray}
\end{widetext}
whose gauge sector is set by (\ref{3rdp}). The disformal metric
\begin{eqnarray}
\label{q-tensor}
Q_{\mu\nu} &=& \left(\frac{1}{16 \pi G_N} +   \frac{c_S}{4} S^2 - \frac{c_O}{8} g^{\alpha\beta} {\mathbb{R}}_{\alpha\beta}(\Gamma)\right) g_{\mu\nu}\nonumber\\ &+& c_{V} {\mbox{tr}}\!\left[V_{\mu}V_{\nu}\right]
\end{eqnarray}
involves the scalars $S$, gauge bosons $V_\mu$ and the affine Ricci curvature $R_{\mu\nu}(\Gamma)$ itself. Here, the Newton's constant
\begin{eqnarray}
G_N^{-1} = 4\pi \sum_m c_m m^2 \xrightarrow{\rm 1-loop} 4 \pi\, {\rm str}[m^2]
\end{eqnarray}
is a trace over mass-squared $m^2$ of all the QFT fields. (As already discussed in \cite{demir1,demir2,demir3},   masses of the known fields (heaviest being the top quark) cannot generate $G_N$ correctly, and this gives a strong case for the necessity of new particles beyond the known ones.) 

The affine curvature sector (\ref{action-affine-2px}) remains stationary against  variations in  $\Gamma^\lambda_{\mu\nu}$ provided that $\Gamma^\lambda_{\mu\nu}$ satisfies the equation of motion
\begin{eqnarray}
\label{gamma-eom}
{}^{\Gamma}\nabla_{\lambda} Q_{\mu\nu} = 0
\end{eqnarray}
where ${}^{\Gamma}\nabla_{\lambda}$ designates the covariant derivative with respect to $\Gamma^\lambda_{\mu\nu}$. This equation of  motion (\ref{gamma-eom}) possesses the general solution
\begin{eqnarray}
\Gamma^\lambda_{\mu\nu} &=& \frac{1}{2} (Q^{-1})^{\lambda\rho}\! \left( \partial_\mu Q_{\nu\rho} + \partial_\nu Q_{\rho\mu} - \partial_\rho Q_{\mu\nu}\right)\\
&=& {}^g\Gamma^\lambda_{\mu\nu} + \frac{1}{2} (Q^{-1})^{\lambda\rho}\! \left( \nabla_\mu Q_{\nu\rho} + \nabla_\nu Q_{\rho\mu} - \nabla_\rho Q_{\mu\nu}\right)\nonumber
\label{aC}
\end{eqnarray}
after using the definition of the Levi-Civita connection in (\ref{LC}). Enormity of the gravitational scale $1/G_N$
enables the affine connection in (\ref{aC}) to be expanded as
\begin{eqnarray}
\Gamma^{\lambda}_{\mu\nu}={}^{g}\Gamma^{\lambda}_{\mu\nu} + 8\pi G_N \left( \nabla_\mu Q^\lambda_\nu + \nabla_\nu Q^\lambda_\mu - \nabla^\lambda Q_{\mu\nu}\right) 
\label{expand-conn}
\end{eqnarray}
up to ${\mathcal{O}}\left(G_N^2\right)$ quadruply Planck-suppressed terms. The use of this affine connection in the affine Ricci tensor in (\ref{a-Ricci}) leads to 
\begin{eqnarray}
{\mathbb{R}}_{\mu\nu}(\Gamma) = R_{\mu\nu}({}^{g}\Gamma) + \ {\mathcal{O}}\left(G_N\right)
\label{expand-curv}
\end{eqnarray}
in which higher-order terms involve derivatives of the scalars $S$ and $V_{\mu}$, with no possibility of inducing any mass term for these fields. 

Replacement of the solution of the affine Ricci curvature in (\ref{expand-curv}) in the gauge boson mass action in  (\ref{3rdp}) in Appendix A leads to the suppression \cite{demir1,demir2,demir3}
\begin{eqnarray}
S_{V}\left(g, {\mathbb{R}}\right) =
 \int d^{4}x  \sqrt{-g} c_V \left\{ 0 + {\mathcal{O}}(G_N) \right\}
\label{3rdpp}
\end{eqnarray}
which restores gauge symmetries up to doubly-Planck suppressed terms. The ${\mathcal{O}}(G_N)$ remainder here does not involve any mass terms for scalars and gauge bosons. 

Having the suppression in (\ref{3rdpp}) at hand, replacement of the affine Ricci curvature in (\ref{expand-curv}) in the total curvature sector in  (\ref{action-affine-2px}) leads to the gravitational action \cite{demir1,demir2,demir3,spurion3}
\begin{eqnarray}
S_{grav}=\int d^4x \sqrt{-g}\!\left\{
\frac{R}{16\pi G_N} +\frac{c_S}{4} S^2 R - \frac{c_O}{16} R^2\right\}
\label{action-grav}
\end{eqnarray}
up to a ${\mathcal{O}}(G_N)$ remainder. Here, $R=g^{\mu\nu} R_{\mu\nu}({}^{g}\Gamma)$ is the Ricci scalar. This is a quadratic-curvature gravity theory in which all the constants $M_{Pl}^2$, $c_S$, $c_O$ are loop-induced constants computed in the flat spacetime QFT. The $R^2$ term, with
\begin{eqnarray}
\label{cO}
c_O =  \frac{(n_B-n_F)}{64 \pi^2} 
\end{eqnarray}
at one loop, is known to realize the Strobinsky inflation \cite{irfan}. This emergent gravity theory differs from the effective action computed in curved spacetime \cite{sakharov,visser} in terms of the origins and values of the parameters. 

As summarized by (\ref{action-grav}) above, gravity has emerged in a way restoring gauge symmetries. In other words, we have obtained gauge symmetry-restoring emergent gravity or briefly {\it symmergent gravity} \cite{demir1,demir2,demir3}. It is clear that the power-law corrections in the flat spacetime effective action in (\ref{power-law}) have left their place to the symmergent gravity action (\ref{action-grav}). This means that the power-law UV sensitivities of the scalar masses, vacuum energy and of the gauge boson masses have all been neutralized or deactivated. Symmergence has thus led to a renormalized QFT in the curved spacetime of (\ref{action-grav}) with gauge hierarchy problem \cite{ghp1,ghp2,ghp3} as well as the charge and color breaking problem \cite{ccb1,ccb2} are resolved, and the gigantic quartic contributions to the cosmological constant have been neutralized \cite{ccp1,ccp2}. It is with this naturalization power that the symmergent gravity differs from all the other emergent gravity theories in  literature (including the Sakharov's induced gravity \cite{sakharov,visser}).

\subsection{Symmergent Gravity as $f(R)$ Gravity}
The symmergent gravity action in (\ref{action-grav}) can be recast as an $f(R)$ gravity action \cite{f(R)review1, f(R)review2,f(R)review3} with
\begin{eqnarray}
\label{fR}
f(R)= R - \pi G_N c_O R^2 
\end{eqnarray}
where $c_O$ is defined in (\ref{cO}).  The resulting field equation 
\begin{eqnarray} 
\label{f1}
\mathit{R}_{\mu \nu} \mathit{F(R)}-\frac{1}{2}g_{\mu \nu}\mathit{f(R)}+[g_{\mu \nu}\Box -\nabla_\mu \nabla_\nu]\mathit{F(R)}= 0
\end{eqnarray}
involves the d'Alembertian operator $\Box$ as well as the  derivative of $f(R)$
\begin{eqnarray}
\label{FR}
F(R) = \frac{d f(R)}{d R} = 1 - 2 \pi G_N c_O R
\end{eqnarray}
Now, trading $f(R)$ for $F(R)$ using the trace of the field equation (\ref{f1}) one is led to the traceless field equation 
\begin{eqnarray} 
\label{f3ss}
\!\!\mathit{R}_{\mu \nu} \mathit{F(R)}-\frac{g_{\mu \nu}}{4}\mathit{ R}\mathit{F(R)}+\frac{g_{\mu \nu}}{4}\Box\mathit{F(R)} -\nabla_\mu \nabla_\nu\mathit{F(R)}\!=\!0
\end{eqnarray}
for $F(R)$ in (\ref{FR}). The solution of this field equation varies with the loop factor $c_O$ and encodes therefore effects of the underlying QFT.

\section{Spherically Symmetric BH Solutions}
In this section we solve gravitational field equations (\ref{f3ss}) for the spherically-symmetric static metric  
\begin{eqnarray} \label{met12}
ds^2=-B(r)dt^2+\frac{dr^2}{B(r)}+r^2(d\theta^2+\sin^2\theta d\phi^2)\,  \end{eqnarray}
parametrized by the radial coordinate $r$, angular coordinates $\theta$ and $\phi$, and yet-to-be determined lapse function $B(r)$. For this metric,  we find
\begin{equation} 
\label{scalar-curve} 
R={\frac { -B'' {r}^{2}-4\,r B'
 -2\,B  +2}{{r}^{2}}}
\end{equation}
for the scalar curvature, and 
\begin{flalign}
\label{E1n}
2FB'' + 2B'F'
- 2BF'' - \frac{4}{r}BF' + \frac{4}{r^2}F (1 - B) &=0, \\
\label{E2n}
2FB'' + 2B'F'
 + 6BF'' -\frac{4}{r}B F' + \frac{4}{r^2}F (1 - B) &=0, \\
\label{E3n}
2FB'' + 2B'F'+ 2BF'' - \frac{4}{r}BF' + \frac{4}{r^2}F (1 - B) &=0 
\end{flalign}
for the equations of motion in (\ref{f3ss}). Here, subtracting any two equations of (\ref{E1n}), (\ref{E2n}) and (\ref{E3n}) from each other yields $B F''  = 0$. Thus it leads consistently  to the linear solution
%for the equations of motion in (\ref{f3ss}). Here, subtracting (\ref{E3n}) from (\ref{E1n}) gives $B F''  = 0$, and subtracting (\ref{E2n}) from (\ref{E1n}) gives again $B F''=0$. These two equations lead thus consistently  to the linear solution
\begin{eqnarray}
\label{FRsoln0}
F[R(r)] = a + b r 
\end{eqnarray}
with $a$ and $b$ undetermined constants.  Equality of the solution (\ref{FRsoln0}) to the definition $F[R(r)]$ in (\ref{FR}) results in the differential equation
\begin{eqnarray}
{\frac { -B'' {r}^{2}-4\,r B'
 -2\,B  +2}{{r}^{2}}}+{\frac {br+a-1}{2\pi\,G_N{c_O}}}=0
\end{eqnarray}
after using the expression of the scalar curvature in (\ref{scalar-curve}). This equation becomes consistent with (\ref{E1n}), (\ref{E2n}) and (\ref{E3n}) only if $b=0$ so that
\begin{eqnarray}
\label{FRsoln}
F[R(r)] = a  
\end{eqnarray}
so that the function $B(r)$ assumes the solution 
\begin{eqnarray}
\label{newmet}
B(r)=1+{\frac {{C}}{r}}+{\frac{ r^2(a-1)}{24 \pi G_N\,c_O}}
\end{eqnarray}
with the additional integration constant $C$. This solution reduces to the usual Schwarzschild solution as  $a\rightarrow 1$ or  $c_O \rightarrow \infty $ provided that constant $C=-2 G_N M$ where $M$ is the total mass within the spherically symmetric 
mass distribution around the origin. It gives AdS/dS BH when $C=0$ (with effective cosmological constant $\Lambda=\frac{1-a}{8 \pi  c_O G_N}$).

	\begin{table}[ht!]
    \centering
    \begin{tabular}{ |p{2cm}||p{1cm}|p{2cm}|p{2cm}|  }
    \hline
        AdS/dS & $\Lambda$ & $a$ & $c_0$ \\ [0.6ex] 
        \hline
        AdS & - & $a>1$ & $c_O>0$ \\
        dS & + & $a>1$ & $c_O<0$ \\
        dS & + & $0<a<1$ & $c_O>0$ \\
        AdS & - &$0<a<1$ & $c_O<0$  \\
       \hline
    \end{tabular}
    \caption{AdS/dS behaviour of the spacetime according to parameters $a$ and loop factor $c_0$.}
    \label{tab:ADStable}
\end{table}

Alternatively, as seen in Table \ref{tab:ADStable} it gives Schwarzschild-AdS BH  when $C=-2 G_N M$ (with effective cosmological constant $\Lambda=\frac{1-a}{8 \pi  c_O G_N}$ for value of $a>1$ and positive value of loop factor $c_0>0$ or for value of $0<a<1$ and negative value of loop factor $c_0<0$ ), on the other hand it gives Kottler (or Schwarzschild-deSitter) BH when $C=-2 G_N M$ and  for value of $0<a<1$ and loop factor $c_0>0$ or for value of $a>1$ and negative value of loop factor $c_0<0$).

\begin{figure}[ht!]
   \centering
    \includegraphics[scale=0.6]{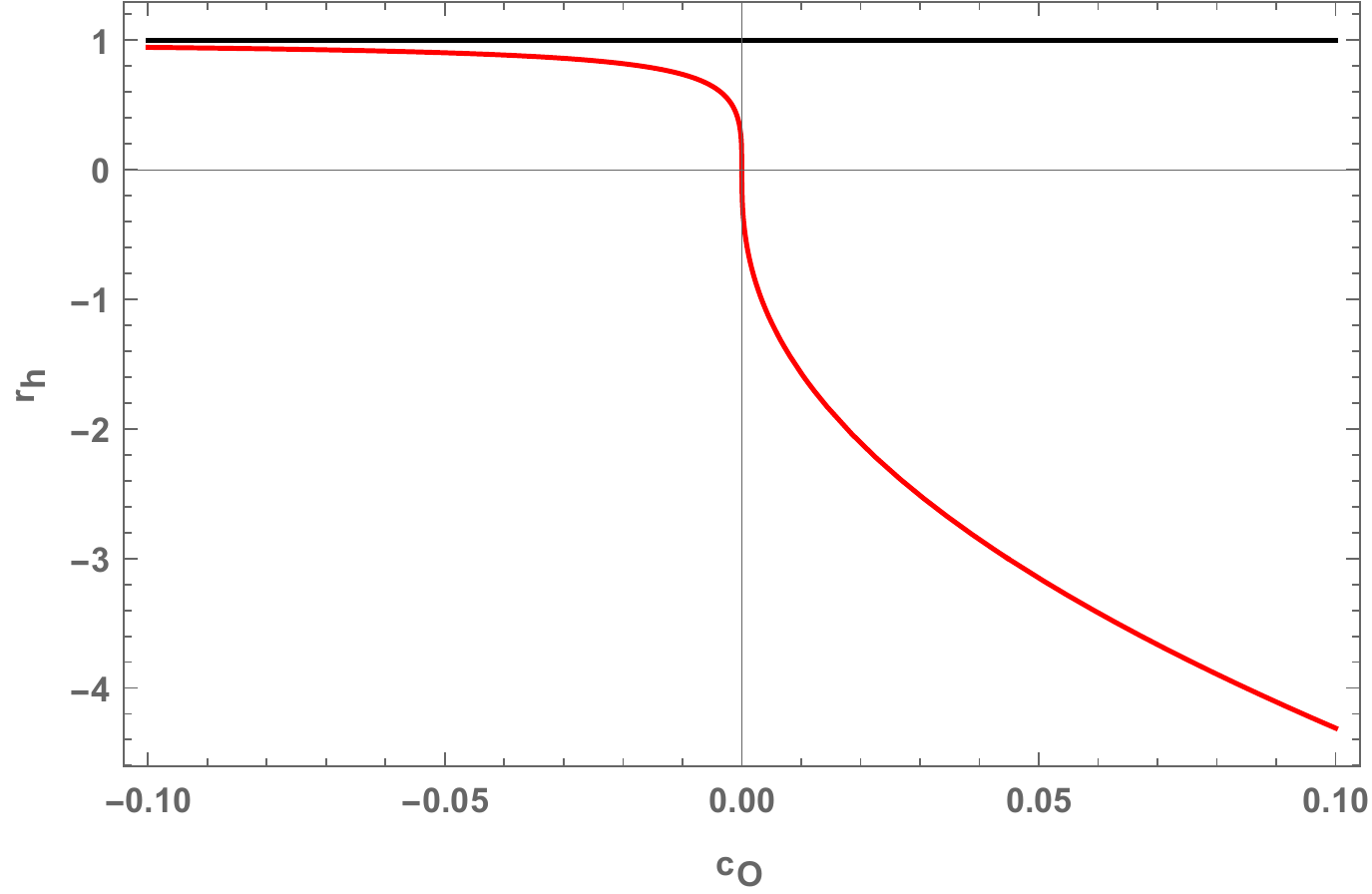}
    \caption{Dependence of the horizon $r_{h}$ on the quadratic-curvature coefficient $c_O$. The symmergent gravity solution (SG, the red curve) nears the Schwarzschild solution (SC, the black curve) for negative $c_O$ (when the underlying QFT has more bosons than fermions) but deviates from it significantly for positive $c_O$ (when the underlying QFT has more fermions than bosons).}
    \label{fig:brvsc}
\end{figure}

In general, the radii $r=r_{h}$ at which largest root of $B(r_{h})=0$ gives the event horizon:
\begin{eqnarray}
\label{horizon}
r_{h}= \frac{h}{(18)^{1/3}} -\frac{(18)^{1/3} c_1}{3 h} 
\end{eqnarray}
where
\begin{eqnarray}
h = ((12 c_1^3 + 81 c_2^2)^{1/2} - 9 c_2)^{1/3}
\end{eqnarray}
with $c_1=1/A$, $c_2=C/A$, $A=(a-1)/(24\pi G_N c_O)$. In spite of its complicated functional form, the BH mass $M$ can be expressed in terms of the horizon $r_{h}$. Depicted in Fig. \ref{fig:brvsc} is the dependence of $r_{h}$ on $c_O$. As is seen from the figure, symmergent gravity solution (SG, the red curve) approaches to the Schwarzschild solution (SC, the black curve) for negative $c_O$ (when the underlying QFT has more fermions than bosons) but deviates from it significantly for positive $c_O$ (when the underlying QFT has more bosons than fermions).

\section{Physical Properties of the BH Solutions}
In general, BHs possess discriminating features which characterize the gravitational theory underlying them \cite{Nashed:2020mnp, Halilsoy:2015qta}.  In the symmergent gravity setup in (\ref{action-grav}), the solution in (\ref{newmet}) with the horizon (\ref{horizon}) is able to probe the model parameter $c_O$. In fact, as defined in (\ref{cO}), $c_O$ is  proportional to the number difference between the bosons and fermions in the underlying QFT, and BH probes actually the QFT's departure from the fermion-boson balance. 

\begin{figure}[ht!]
   \centering
   \includegraphics[scale=0.6]{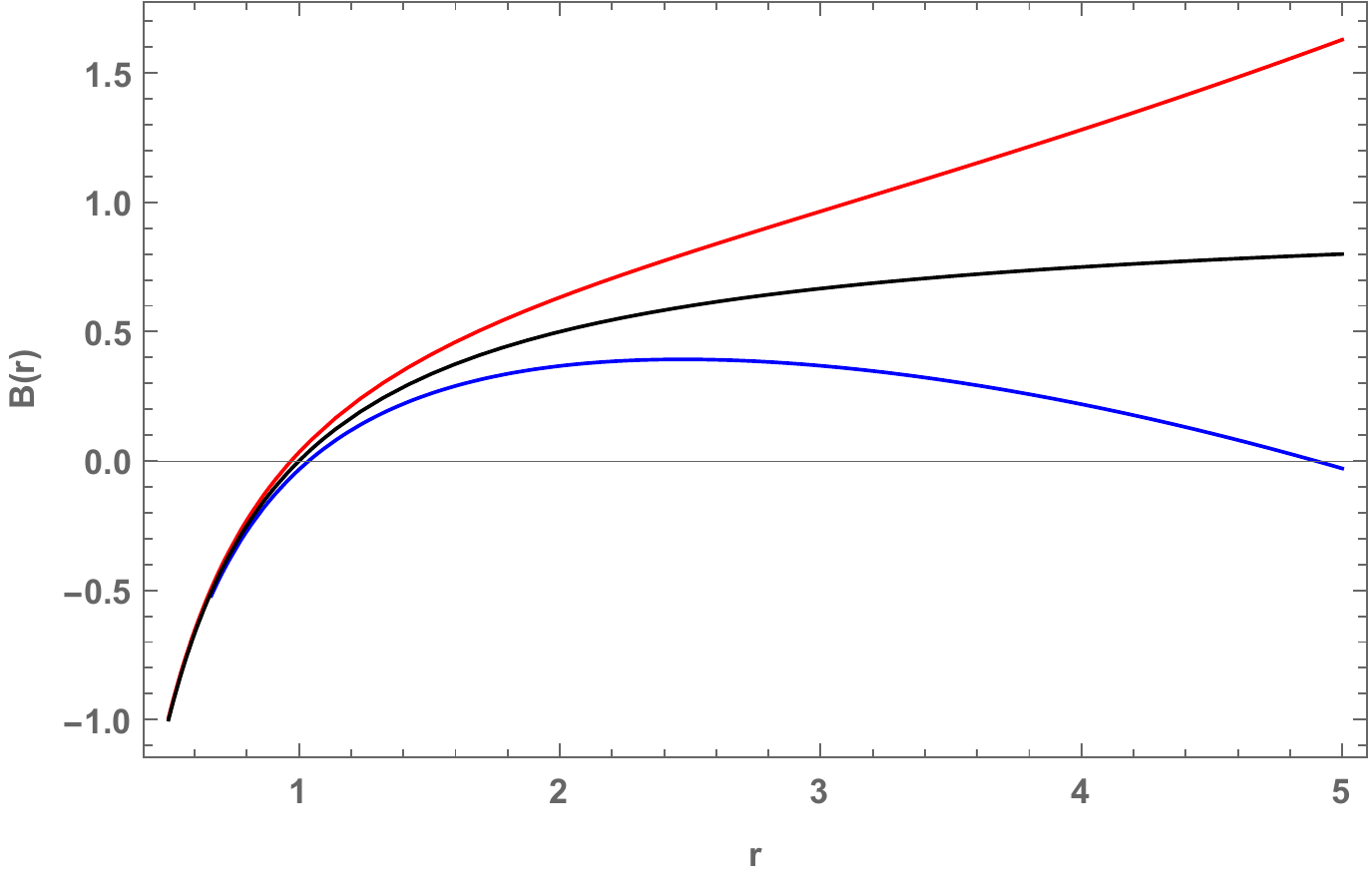}
    \caption{The lapse function $B(r)$ as a function of $r$ for $M=a=0.5$ ($M$ in gravitational units) and for Schwarzschild solution (SC, the black curve) and for the symmergent gravity with loop factor $c_O=+0.2$ (blue curve) and $c_O=-0.2$ (red curve). The zeros of these curves give the event horizon of the corresponding black hole.}
    \label{fig:brvsr}
\end{figure}

As shown in Fig.\ref{fig:brvsr}, sign of $c_O$ directly affects the lapse function $B(r)$. If the underlying QFT has more (less) bosons than fermions then $B(r)$ increases (decreases) with $r$.  
In Fig.\ref{fig:brvsr}, each $B(r)$ intersects the horizontal  axis  at the radius giving its horizon. Here, positive value $c_O=0.2$ (blue curve) generates two  horizons: the inner horizon ($r=r_-$) and outer horizon ($r=r_+$). Negative value $c_O=-0.2$ (red curve) leads to one horizon. Black curve represents the Schwarzschild horizon for $r_h=1$. It is worth emphasizing that the metric (\ref{met12}) is not asymptotically  flat since
$\lim _{r \rightarrow \infty} g_{t t}=\lim _{r \rightarrow \infty} g^{r r}$.

The horizon corresponds to a removable singularity. The essential singularity occurs at the point where the Kretschmann scalar
\begin{eqnarray}
R_{\alpha \beta \delta \gamma} R^{\alpha \beta \delta \gamma}&=&{\frac { B''^{2}{r}^{4}+4B'^{2}{r}^{2}+4\, \left( B -1 \right) ^{2}}{{r}^{4}}}\nonumber\\
&=&\frac{48 G_N^2 M^2}{r^6} + \frac{(a-1)^2}{24 \pi G_N^2 c_O^2 }
%\frac{1152 c_O^2 G_N^4 M^2 \pi^2 +(a-1)^2 r^6}{24c_O^2 G_N^2 \pi^2 r^6}
\end{eqnarray}
diverges and the said singular point, $r=0$, occurs because of the Schwarzschild part not the quadratic curvature part with the  coefficient $c_O$.

\begin{figure}[ht!]
   \centering
    \includegraphics[scale=0.6]{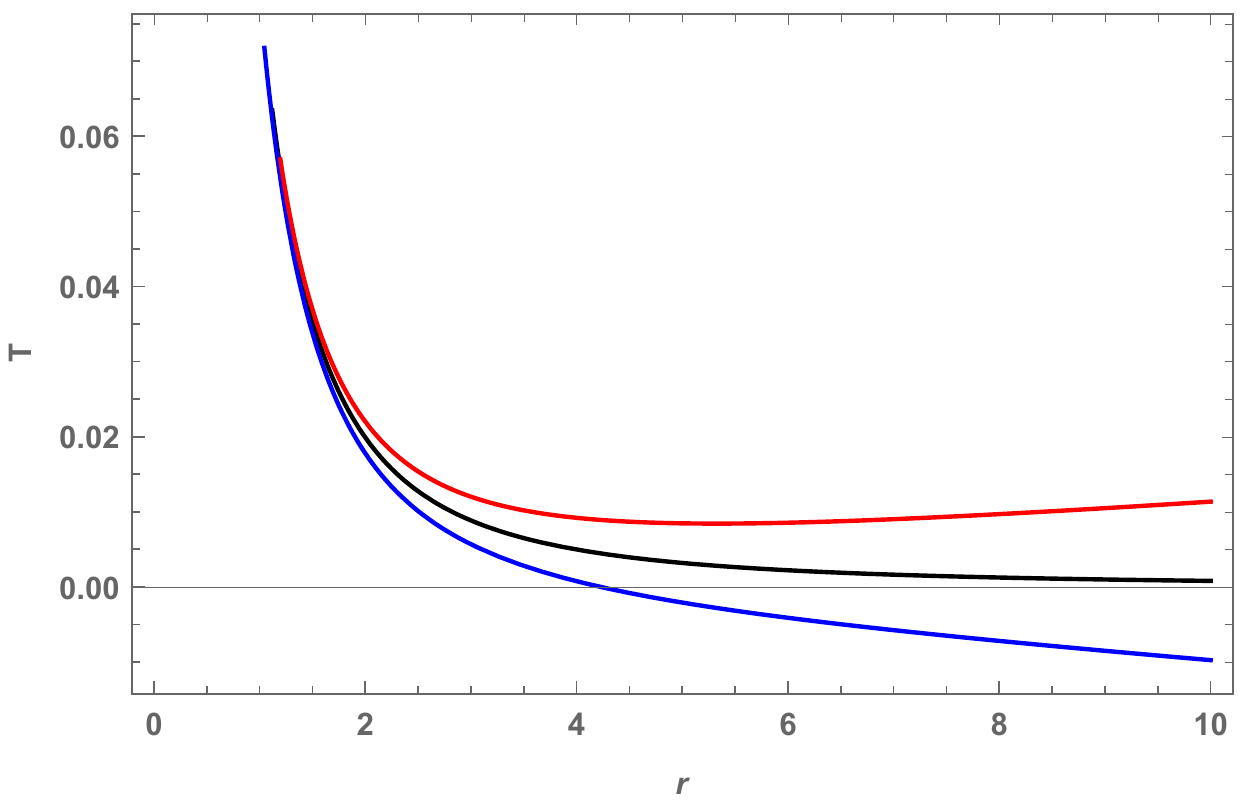}
    \caption{Hawking temperature $T$ versus $r=r_h$ for the Schwarzschild (black curve), $c_O=1$ (blue curve) and $c_O=-1$ (red curve) at $a=0.5$ and $2 G_N M =1$.}
    \label{fig:tempvsc}
\end{figure}

\begin{figure}[ht!]
   \centering
    \includegraphics[scale=0.6]{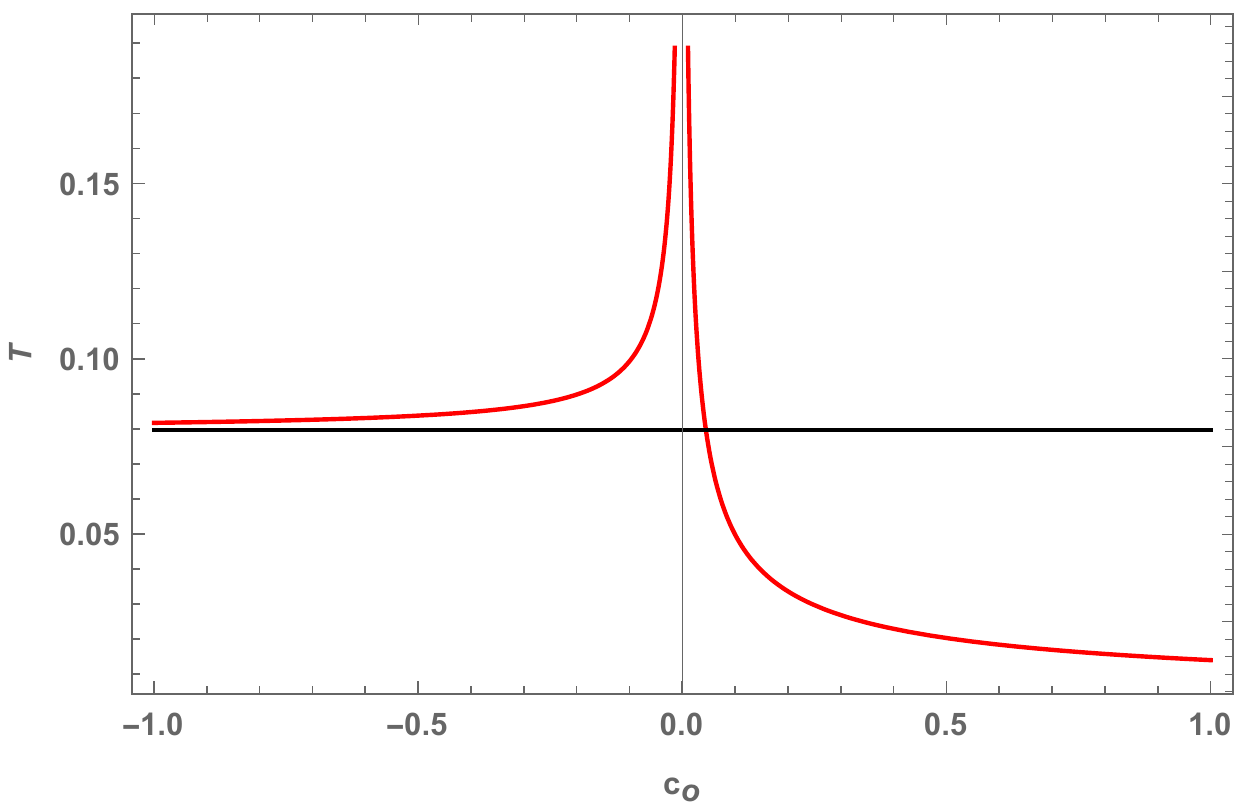}
    \caption{Hawking temperature $T$ versus $c_O$ for $a=0.5$ and $2 G_N M =1$. Black line represents the Schwarzschild case.}
    \label{fig:temp}
\end{figure}

The Hawking temperature $T$ of a BH follows from its surface gravity at the location of the horizon \cite{Hawking:1974sw}
\begin{equation}
k_{h}^{2}=-\frac{1}{2} \lim _{r \rightarrow r_{h}}\left(D_{\mu} \bar{K}_{v}\right)\left(D^{\mu} \bar{K}^{v}\right)
\end{equation}
where $D^{\mu}$ is covariant derivative with respect to the metric (\ref{met12}), and  $\bar{K}$ is the  timelike Killing vector with normalisation constant $\gamma_t$  (with $K=\gamma_{t} \frac{\partial}{\partial t}$). For a spherically symmetric BH the surface gravity is
\begin{equation}
k_{h}=\frac{1}{2} \frac{1}{\sqrt{-g_{t t} g_{r r}}}\left|g_{t t, r}\right|_{r=r_{h}}
\end{equation}
and hence, the Hawking temperature of the symmergent gravity BH turns out to be 
\begin{eqnarray}
T=\frac{k_{h}}{2 \pi} =\frac{G_N M}{8 r_{h}^2}+\frac{(1-a)r_{h}}{192\pi c_O G_N}
\end{eqnarray}
whose variation with $r_h$ is plotted in Fig. \ref{fig:tempvsc} and variation with loop factor $c_0$ is plotted in Fig. \ref{fig:temp}. As seen from the plot, Hawking temperature increases (decreases) with $r_+$ for $c_O=-1$ ($c_O=1$).

\begin{figure}[ht!]
   \centering
    \includegraphics[scale=0.6]{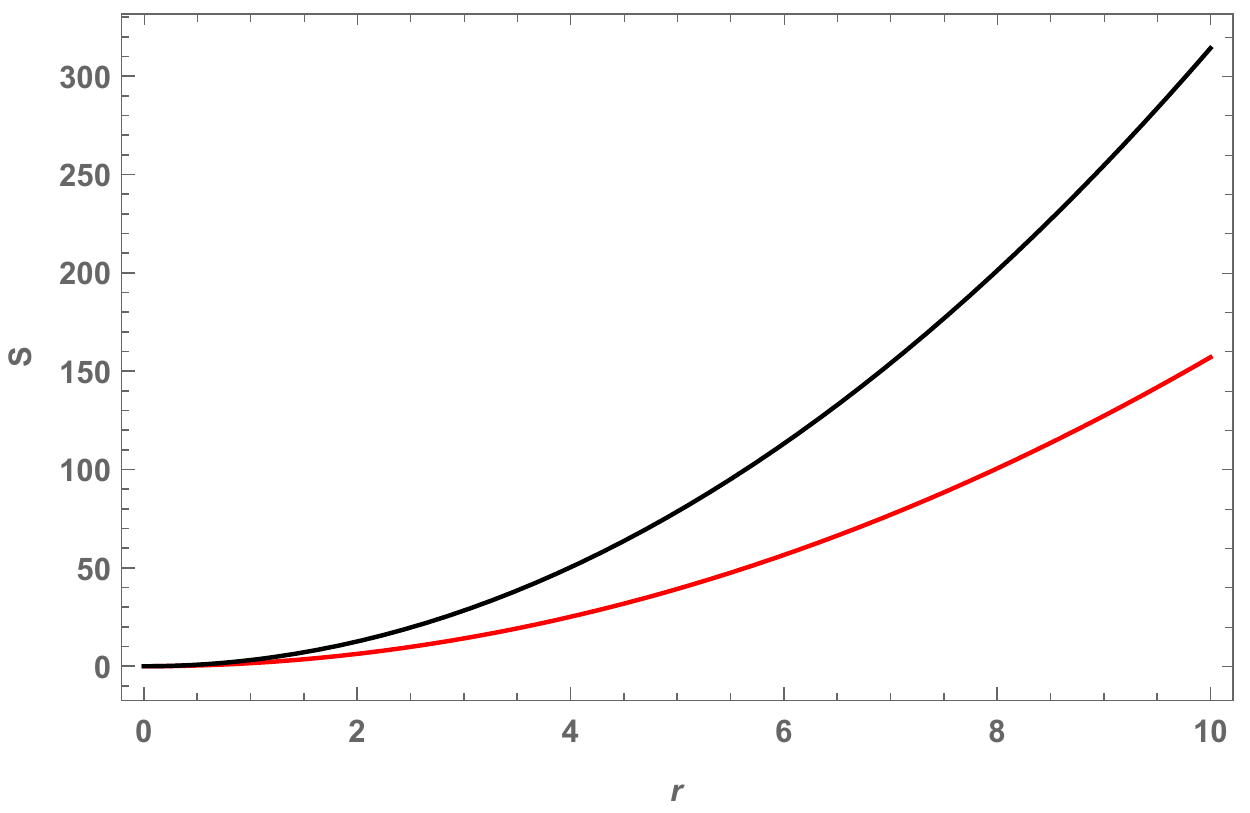}
    \caption{Bekenstein-Hawking entropy $S$ versus $r=r_h$ for the Schwarzschild (black curve) and symmergent gravity (red curve) BH solutions for loop factor $c_O=1$  and $2 G_N M =1$.}
    \label{fig:entropy}
\end{figure}

\begin{figure}[ht!]
   \centering
    \includegraphics[scale=0.6]{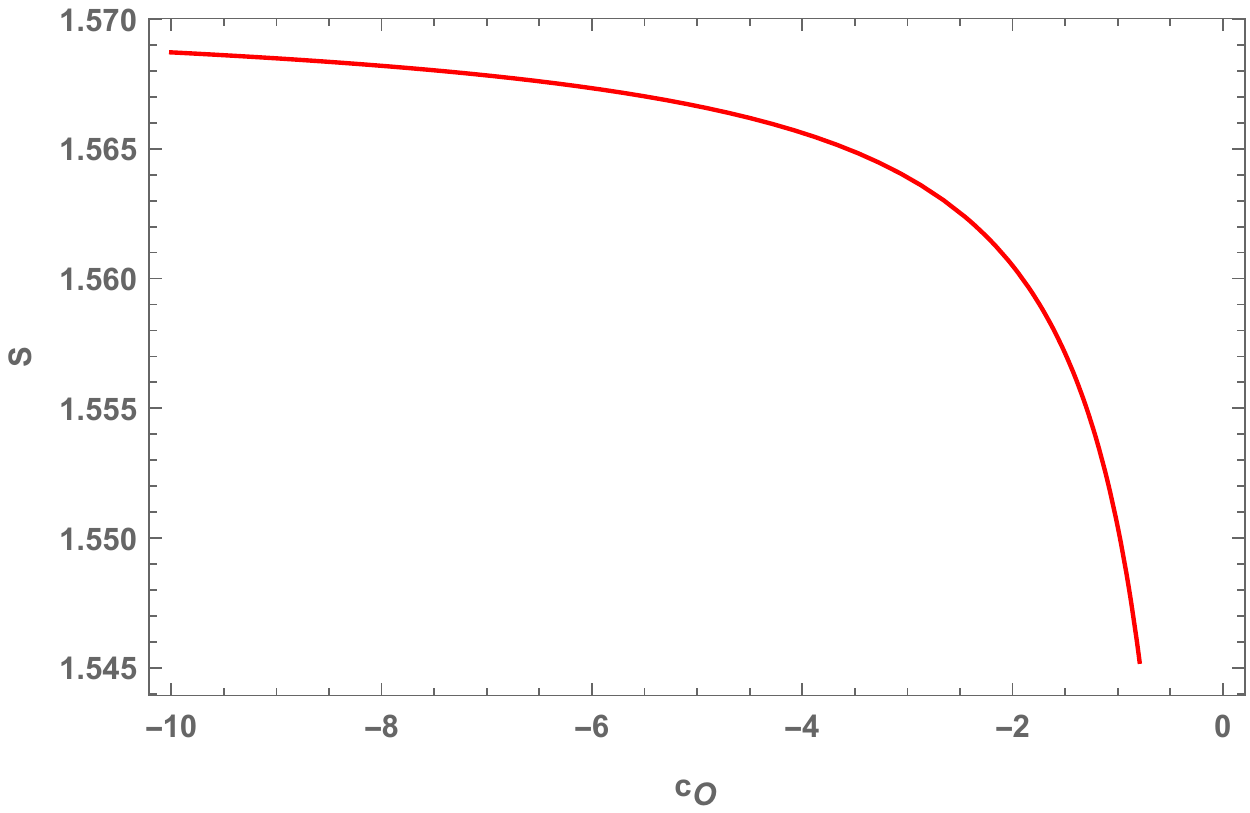}
    \caption{The Bekenstein-Hawking entropy $S$ versus $c_O$ at $a=0.5$ and $2 G_N M =1$.}
    \label{fig:entropy2}
\end{figure}

The Bekenstein-Hawking entropy \cite{Bekenstein:1973ur,odintsov,Gong:2007md,Brustein:2007jj} is given by 
\begin{equation}
S\left(r_{h}\right)=\frac{1}{4 G_{N}} \mathcal{A}_h(r_h) F\left(r_{h}\right)
\end{equation}
where $\mathcal{A}_{h}= 4 \pi r_{h}^{2}$ is the horizon area, and $F(r_h)=a$ as found in (\ref{FRsoln}) so that the entropy takes the form
\begin{equation}
S=\frac{a \pi r_{h}^2}{G_N}
\end{equation}
which is independent of $c_O$. Its variation with $r_h$ is depicted in Fig. \ref{fig:entropy}, where it is seen that growth of the entropy with $r_h$ is controlled by the parameter $a$. Entropy remains positive, which is the physical result. The variation of entropy with loop factor $c_0$ is plotted in Fig. \ref{fig:entropy2}.

\section{Black Hole Shadow in Symmergent Gravity}

To calculate the shadow of the black hole in symmergent gravity, we first obtain the equations of null geodesics and write the Hamiltonian of the moving photon:
\begin{equation}\label{hamij}
2H=g^{ij}p_{i}p_{j}=0,
\end{equation}
Because of the spherically symmetry, in the equatorial plane with $\theta=\pi/2$, the above equation Eq. (\ref{hamij}) is written as
\begin{equation}\label{ramo}
\frac{1}{2}\left[-\frac{p_{t}^{2}}{B(r)}+B(r)p_{r}^{2}+\frac{p^{2}_{\phi}}{r^{2}}\right]=0.
\end{equation}
The quantities of constants motion $p_t$ and $p_{\phi}$ which are related to energy  $-p_t=E$ and angular momentum of the photon $p_{\phi}=L$ can be found as $ \dot{p}_{t}=-\frac{\partial H}{\partial t}=0$ and $\dot{p}_{\phi}=-\frac{\partial H}{\partial \phi}=0.$
Then it is straightforward to obtain the equations of motion for the photon:
\begin{equation}
\dot{t}=\frac{\partial H}{\partial p_{t}}=-\frac{p_{t}}{B(r)},
\end{equation}
\begin{equation}
\dot{\phi}=\frac{\partial H}{\partial p_{\phi}}=\frac{p_{\phi}}{r^{2}},
\end{equation}
\begin{equation}
\dot{r}=\frac{\partial H}{\partial p_{r}}=p_{r}B(r),
\end{equation}
with the radial momentum $p_{r}$. Using the above equations, one can define the effective potential of photon using the radial equation of motion as $V_{eff}+\dot{r}^{2}=0$,
where the effective potential is
\begin{equation}\label{e1}
V_{eff}=B(r)\left[\frac{L^{2}}{r^{2}}-\frac{E^{2}}{B(r)}\right].
\end{equation}
Maxima of the effective potential corresponds to unstable circular orbits which are found by taking the maximal value of the effective potential $V_{eff}=V^{\prime}_{eff}=0$ \cite{Chandrasekhar:1985kt}. Using the condition for the turning point $r_{p}$ of a photon, where $\dot{r}=0$ or $V_{eff}=0$, the impact parameter $b$ is

\begin{equation}
b=\frac{L}{E}=\frac{r}{\sqrt{B\left(r\right)}},
\end{equation}

and using $V_{eff}=V^{\prime}_{eff}=0$, gives

\begin{eqnarray}
\frac{B^{\prime}(r)}{B(r)}=\frac{2}{r},
\end{eqnarray}
where its solution gives us the radius of the photon sphere $r_p=3GM$.
On the other hand, using the Eq.  (\ref{ramo}) within the orbit equation $
\frac{dr}{d\phi}=\frac{\dot{r}}{\dot{\phi}}=\frac{r^{2}B(r)p_{r}}{j}$, this relation is found:
\begin{equation}
\frac{dr}{d\phi}=\pm r\sqrt{B(r)\left[\frac{r^{2}E^{2}}{B(r)L^{2}}-1\right]}.
\end{equation}
and the above equation reduces to this form 
\begin{equation}\label{eosf}
\frac{dr}{d\phi}=\pm r\sqrt{B(r)\left[\frac{r^{2}B(R)}{B(r)R^{2}}-1\right]},
\end{equation}
by using the photon orbit at the turning point
\begin{equation}
\left.\frac{dr}{d\phi}\right|_{r=R}=0,
\end{equation}
and
\begin{equation}
\frac{E^{2}}{L^{2}}=\frac{B(R)}{R^{2}}.
\end{equation}

One can assume that light rays are coming from a static observer located at position $r_{o}$ and transmitting into the past with an angular radius $\Theta$ with respect to the radial direction:
\begin{equation}
\cot\Theta=\frac{\sqrt{g_{rr}}}{\sqrt{g_{\phi\phi}}}\cdot\frac{dr}{d\phi}{\Big{|}}_{r=r_{o}}=\frac{1}{r\sqrt{B(r)}}\cdot\frac{dr}{d\phi}{\Big{|}}_{r=r_{o}},
\end{equation}
then the above equation reduces to
\begin{equation}
\cot^{2}\Theta=\frac{r_{o}^{2}B(R)}{B(r_{o})R^{2}}-1,
\end{equation}
which can be rewritten as \begin{equation}
\sin^{2}\Theta=\frac{B(r_{o})R^{2}}{r_{o}^{2}B(R)}.
\end{equation}

By letting $R\to r_{p}$ with $r_{p}$ the circular orbit radius of the photon, the critical impact parameter of BH is \cite{ Lake:1977ui,stuchlik} 

\begin{equation}
b_{\mathrm{cr}}=\frac{r_p}{\sqrt{B(r_p)}},
\end{equation}
is obtained from the photon sphere radius, $r_p$, as the angular semi-diameter of the shadow around the BH as seen by a distant observer. The angular shadow radius for a static observer $r_{0}$ at a large distance is \cite{Perlick:2018iye,Maluf:2020kgf} \begin{equation}\label{R44}
\sin ^{2} \Theta=\frac{B(r_{o}) b_{\mathrm{cr}}^{2}}{r_{o}^{2}}.
\end{equation}

\begin{figure}[ht!]
   \centering
    \includegraphics[scale=0.5]{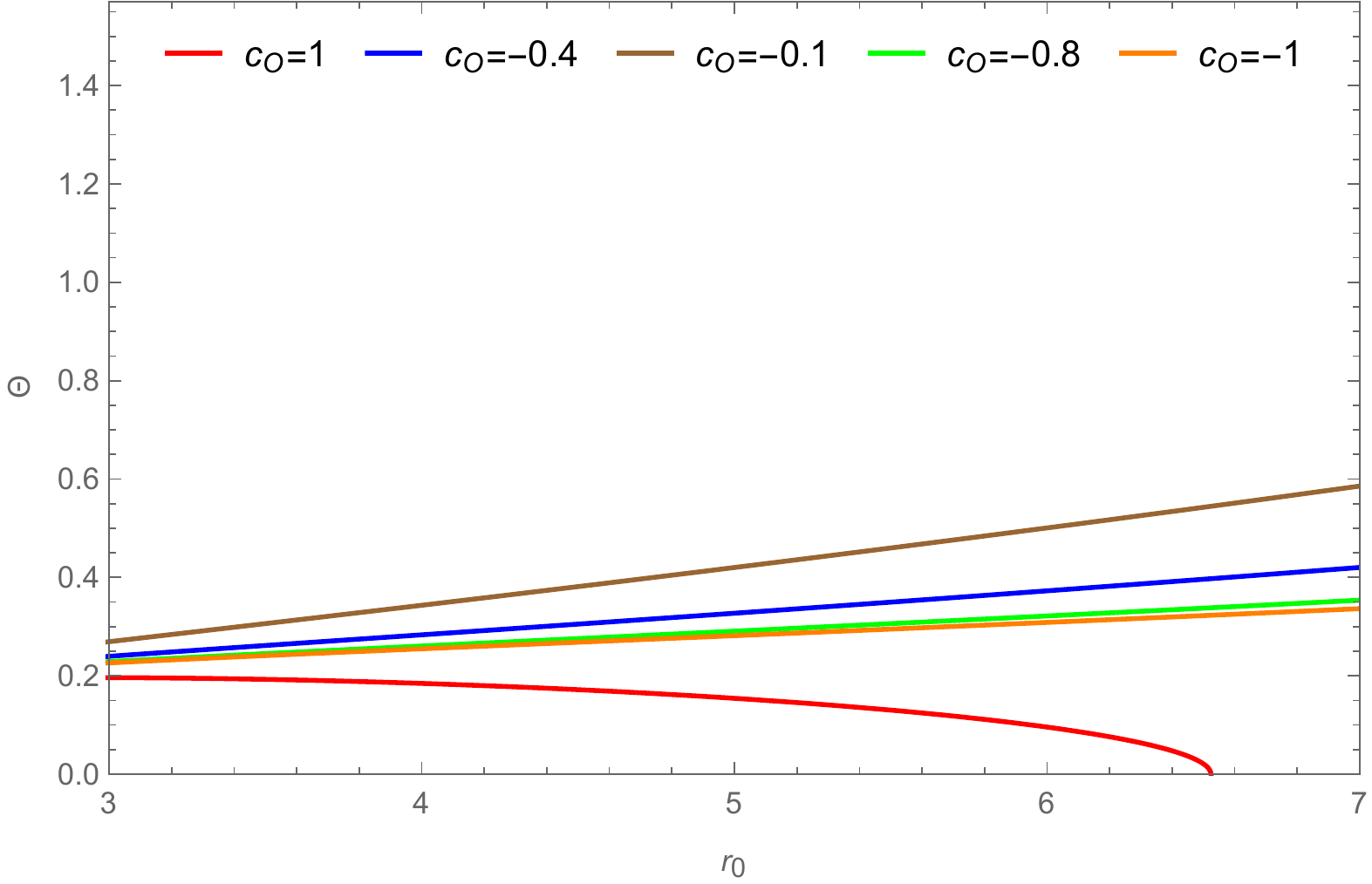}
    \caption{The BH shadow angular radius $\Theta$ seen by a static observer at different locations $r_0$ for various $c_O=1$  (blue), and $c_O=-1$ (red) at fixed values of $a=0.5$ and $2 G_N M=1$. Note that the angular radius become zero, when the observer is located at the cosmological horizon (the sky is completely bright for observer).}
    \label{fig:rovst}
\end{figure}

Depicted in Fig. \ref{fig:rovst} is the angular shadow radius $\Theta$ versus a static observer at different locations $r_0$.
 For asymptotically flat static black holes the shape of the shadow is nothing but the standard circle since photons coming from both sides of the static BH have the same value of the deflection angle, on the other hand, the shadow in the Kottler spacetime which is nonsymptotically flat such a case the shadow is not circular \cite{Perlick:2018iye}.

Fig. \ref{fig:rovst} shows the dependence of the shadow angular radius $\Theta$ on $c_O$. As follows from the Fig. \ref{fig:rovst}, the BH shadow angular radius increase for increasing the negative value of $c_O$, on the other hand BH shadow angular radius decrease for the positive value of loop factor $c_0$. It is clear that at a larger static observer distance $r_0$, shadow angular radius
$\Theta$ depends on the value of the loop factor $c_0$.

\subsection{Deflection Angle using Rindler-Ishak method}
\begin{figure}[ht!]
   \centering
    \includegraphics[scale=0.6]{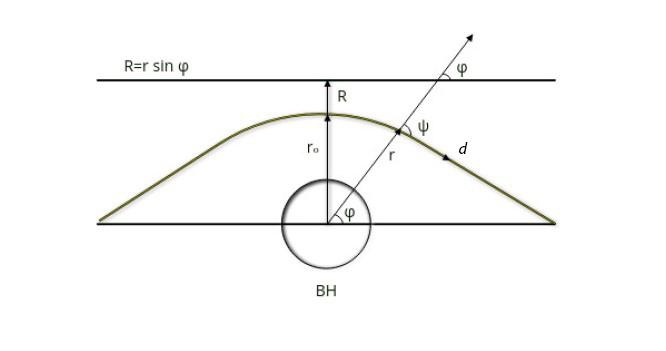}
    \caption{Illustration of gravitational lensing effect}
    \label{fig:graph}
\end{figure}

It is clear that light bends around the BH and also other compact objects \cite{Ovgun:2019wej,Ovgun:2018fnk,Javed:2020lsg,Jusufi:2017uhh,Li:2020dln,Li:2020wvn,Jusufi:2017mav,Javed:2019rrg,Javed:2019ynm,Javed:2019qyg}. As shown in Figure \ref{fig:graph}, the angle $\psi$ between the photon orbit direction $d$ (of components $d^i$) and the direction ${\delta}$ (of components $\delta^i$) along $\varphi=$ constant line is given by the Rindler-Ishak formula \cite{Rindler:2007zz} 
\begin{equation}
\cos\psi=\frac{g_{ij}d^{i}\delta^{j}}{\sqrt{g_{ij}d^{i}d^{j}}\sqrt{g_{ij}\delta^{i}\delta^{j}}},
\label{7}
\end{equation}
where $g_{ij}$ are the coefficients of the 2-metric on $\theta=\frac{\pi}{2}$, $t=$ constant surface. Substituting $d=(dr,d\varphi)$ and $\delta=(\delta r,0)$ in (\ref{7}) one gets
\begin{equation}
\cos\psi=\frac{|dr/d\varphi|}{\sqrt{|dr/d\varphi|^2+B(r)r^2}},
\label{8}
\end{equation}
%or equivalently
%\begin{equation}
%\tan\psi=\frac{r\sqrt{B(r)}}{|dr/d\phi|}.
%\label{9}
%\end{equation}
where to obtain one-sided deflection angle at small $\psi$ ($\psi\ll 1$) we use
\begin{equation}
\frac{1}{r} = \frac{\sin (\varphi)}{R}+\frac{3 m}{2 R^{2}}\left(1+\frac{1}{3} \cos (2 \varphi)\right), \label{10}
\end{equation}
in which the parameter $R$ is an impact parameter related to the distance $r_0$ of closest approach by the formula 
\begin{equation}
\frac{1}{r_{0}}=\frac{1}{R}+\frac{m}{R^{2}},
\end{equation}
so that we get the photon orbit equation 
\begin{eqnarray}
\frac{d r}{d \varphi}=\frac{m r^{2}}{R^{2}} \sin (2 \varphi)-\frac{r^{2}}{R}. 
%\cos (\phi) \equiv A(r, \phi)
\label{11}
\end{eqnarray}
Finally, by substituting this orbit equation in (\ref{8}) the total deflection angle ($2\epsilon=2\psi=\hat{\alpha}$ for $\varphi=0$) takes the form
\begin{eqnarray}
\hat{\alpha} \approx  \frac{4 m}{R}\left(1-\frac{2 m^{2}}{R^{2}}-\frac{(1-a) R^{4}}{8 \pi  c_O G_N 24 m^{2}}\right), \label{d1}
\end{eqnarray}

in which the first term is the usual Schwarzschild result. In
addition to the usual term, the last term with $R^3$, $a$ and loop factor $c_0$ parameters is that of the Schwarzschild-de Sitter spacetime \cite{Rindler:2007zz}.  

\begin{figure}[ht!]
   \centering
    \includegraphics[scale=0.6]{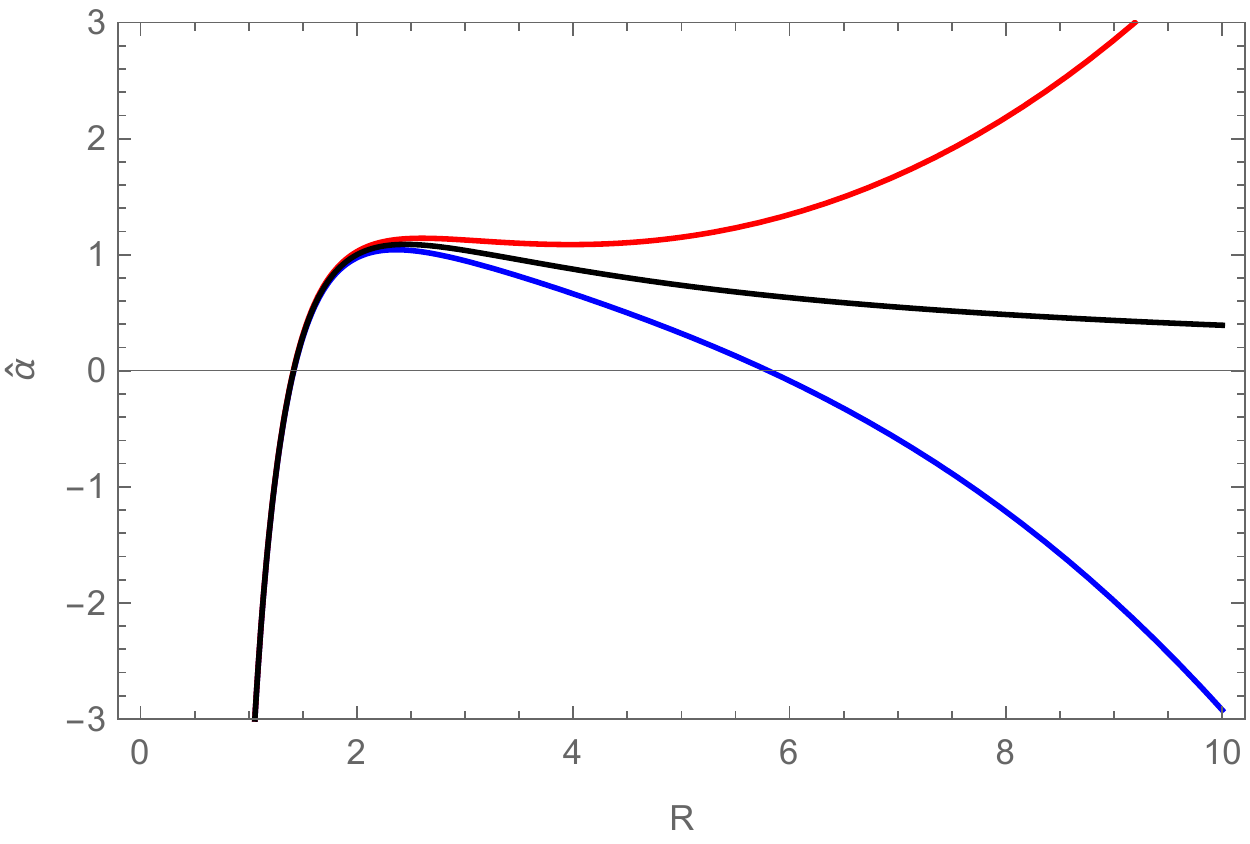}
    \caption{The deflection angle $\hat{\alpha}=2\psi$ versus the radius $R$ for Schwarzschild solution (black curve) and symmergent gravity 
    ($c_O=1$ (blue curve) and $c_O=-1$ (red curve)) at $a=0.5$ and $2 G_N M=1$.}
    \label{fig:defangle}
\end{figure}

Plotted in Fig. \ref{fig:defangle} is the photon deflection angle $\hat{\alpha}=2\psi$ for different $c_O$ values. It is obvious that the positive/ negative parameter $c_O$ decreases/ increases with the weak deflection angle $\hat{\alpha}$ as seen in Fig. \ref{fig:defangle}. Hence, we show that the deviation from general relativity: when loop factor parameter $c_O>0$, the weak deflection angle $\hat{\alpha}$ is smaller than it by Schwarzschild black hole; when $c_O<0$, the weak deflection angle $\hat{\alpha}$ is larger than it by Schwarzschild black hole. The results show that for the increasing impact parameter $R$, weak deflection $\hat{\alpha}$ increases for Schwarzschild-AdS BH, and decreases for the Schwarzschild-dS BHs. 

The dependence of the deflection angle on $c_O$ in this figure reveals that deflection direction gets reversed with the reversal of the $c_O$ sign at large $R$. This sensitivity to the sign of $c_O$ shows that the bending of light around the BH is a sensitive probe of the number difference between the bosons and fermions in the underlying QFT.  

\section{Effects of Plasma on BH Shadow}
The effects of plasma on the BH shadow worth a separate discussion. Effects of nonmagnetized cold plasma with electron plasma frequency \cite{roger}
\begin{equation}
\omega_{p}(r)^{2}=\frac{4 \pi e^{2}}{m_e} N(r),
\end{equation}
on the BH shadow in symmergent gravity (\ref{newmet}) can be studied following the methods of \cite{Atamurotov:2015nra,Belhaj:2020rdb,plasmashadow}. In this frequency formula,  $e$ is electron electric charge, $m_e$ is its mass, and $N(r)$ is the number density of the electrons
in the plasma. The refraction index $n(\omega_p)$ of this plasma is given by \cite{Atamurotov:2015nra}
	\begin{equation}\label{12t}
	n(\omega_p)^2 =1-\Big(\frac{\omega_p(r)}{\omega_0(r)}\Big)^2,
	\end{equation}
where $\omega_0(r)$ is the photon frequency. Considering a radial power-law density  $N(r)=\frac{N_0}{r^s}$ \cite{roger}, this refractive index takes the form 
%	\begin{eqnarray}
%	\omega_p(r)^2 = \frac{k}{r^h}~,~k\geqslant 0.
%	\end{eqnarray}
%	The refractive index $n$ therefore takes the form 
	\begin{equation}
	n(\omega_p)^2=1-\frac{k}{r^s},
	\end{equation} 
where $k=\frac{4\pi e^2 N_0}{m_e \omega_0^2}$. In general, different $s$ values correspond to different plasma properties. For weakest $r$ dependence, it is convenient to take $s=1$ \cite{Abdujabbarov:2015pqp}. 

In the presence of the plasma, photon sphere lies at 
	\begin{equation}\label{25}
	\Bigg(nr B' - 2n B-2n' r B\Bigg)\Bigg\vert_{r=r_{\hat{p}}}=0,
	\end{equation}
with radius of the photon sphere $r=r_{\hat{p}}$ in plasma medium. The angular radius of the BH shadow for a static observer $r_{0}$ at a large distance then takes the form \cite{Belhaj:2020rdb}
\begin{eqnarray} 
sin \Theta^{(pl)}&=\frac{ r_{\hat{p}} \sqrt{B(r_{0})}}{r_{o} n(r_{\hat{p}})(\sqrt{B(r_{\hat{p}})}}
%\\ R_{s}^{2(pl)}&=\frac{5 n(r_{\hat{p}})^{2} r_{\hat{p}}^{2}+2 n(r_{\hat{p}}) n^{\prime}(r_{\hat{p}}) r_{\hat{p}}^{3}B(r_{0})}{n(r_{\hat{p}})^{2}\left(3 B(r_{\hat{p}})+r_{\hat{p}} B^{\prime}(r_{\hat{p}})\right)}\Bigg\vert_{r=r_{\hat{p}}},
\end{eqnarray}

whose dependence on $k$ (sensitivity to plasma). %is shown in Fig \ref{tab:shadowtable}, where it is seen that plasma does not have a strong impact on the BH shadow in the symmergent gravity.
	
\begin{figure}[ht!]
    \centering
    \includegraphics[scale=0.45]{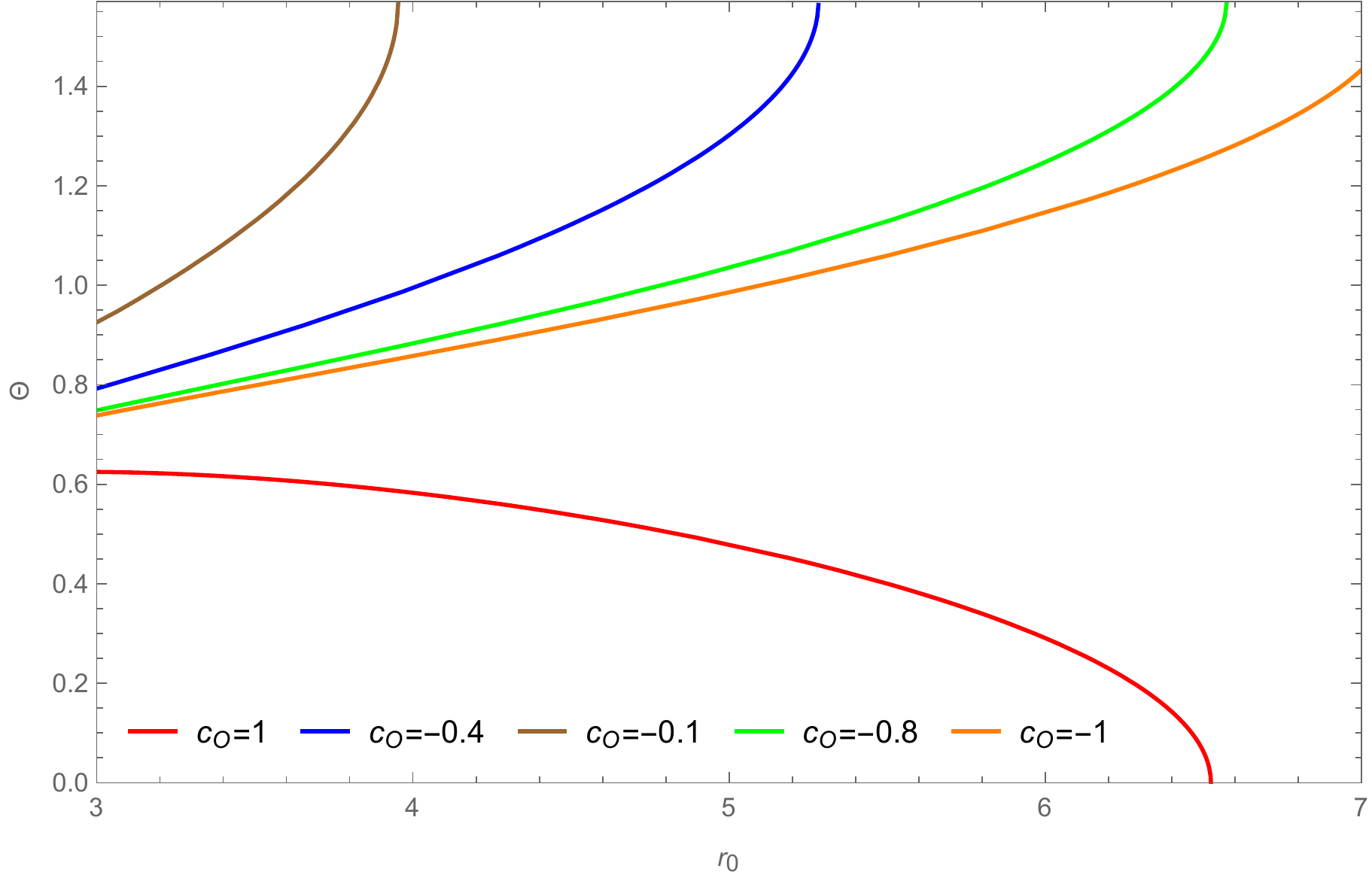}
    \caption{The BH shadow angular radius $\Theta$ seen by a static observer at different locations $r_0$ under plasma medium for various loop factor $c_0$  values at fixed $2 G_N M=1$ and $k=1$.}
    \label{fig:k1}
\end{figure}

\begin{figure}[ht!]
    \centering
    \includegraphics[scale=0.5]{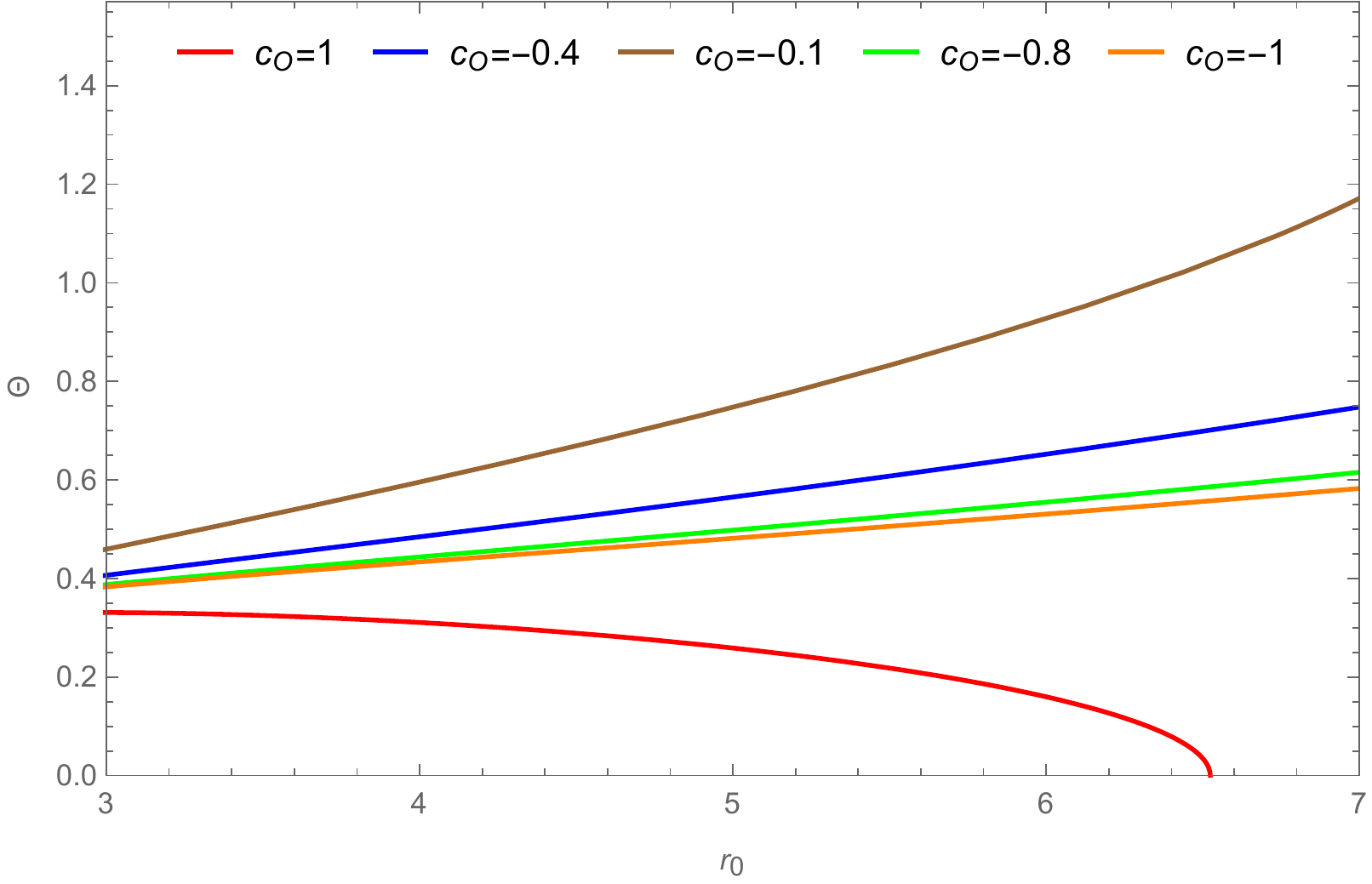}
    \caption{The BH shadow angular radius $\Theta$ seen by a static observer at different locations $r_0$ under plasma medium for various loop factor $c_0$  values at fixed $2 G_N M=1$ and $k=0.6$.}
    \label{fig:k2}
\end{figure}

\begin{figure}[ht!]
    \centering
    \includegraphics[scale=0.5]{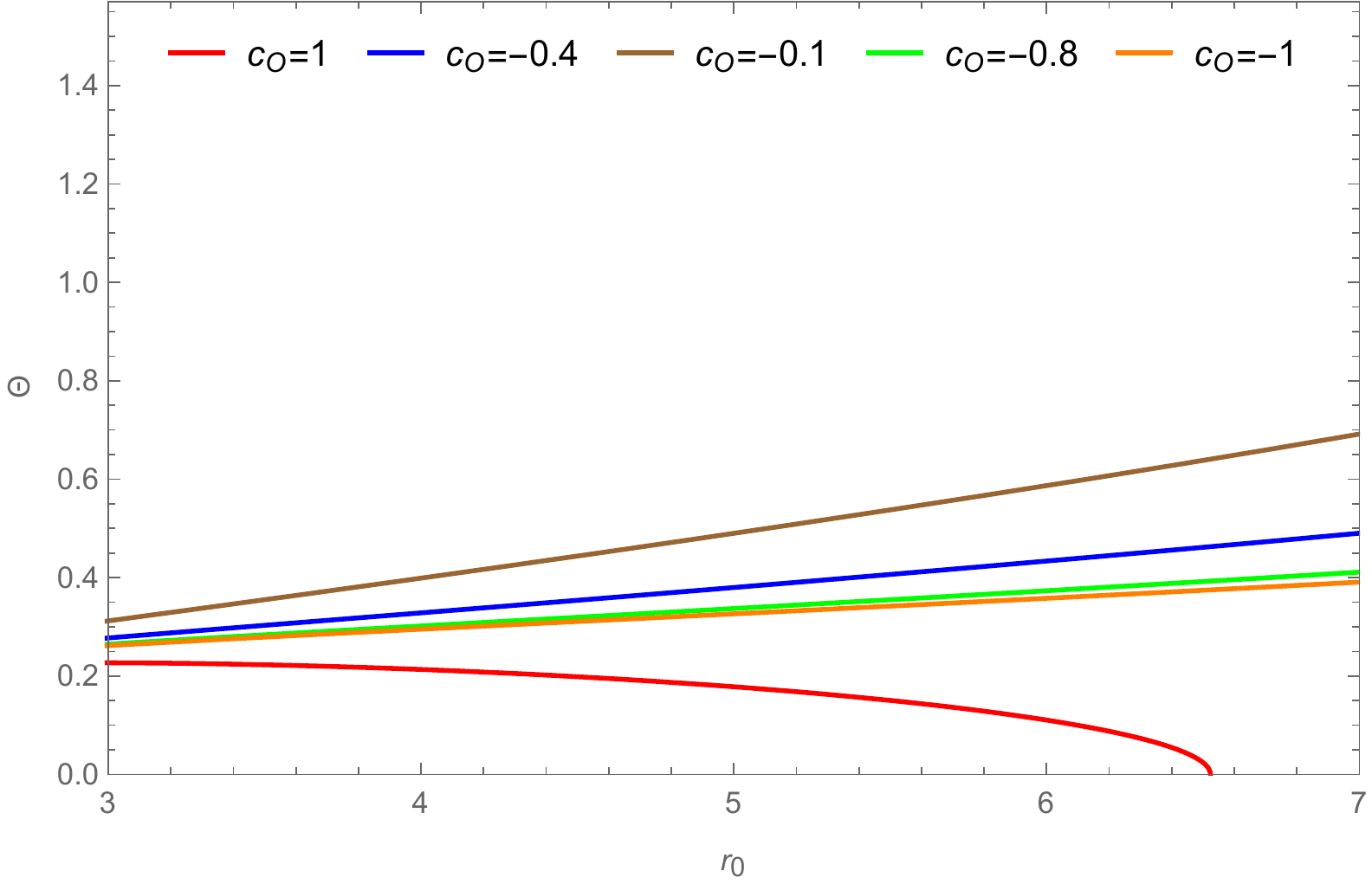}
    \caption{The BH shadow angular radius $\Theta$ seen by a static observer at different locations $r_0$ under plasma medium for various loop factor $c_0$  values at fixed $2 G_N M=1$ and $k=0.2$.}
    \label{fig:k3}
\end{figure}

Effects of the plasma (various $k$ values) are depicted in 
Fig. \ref{fig:k1} for symmergent gravity with $k=1$, in Fig.
\ref{fig:k2} for symmergent gravity with $k=0.6$, and  in Fig. \ref{fig:k3} for symmergent gravity with $k=0.2$. 

 In general, under plasma medium as $k$ decreases, $\Theta$ decreases appreciably for symmergent gravity. It turns out that plasma have a impact on the BH shadow in symmergent gravity.

\section{Conclusion}
In the present work, we have performed a comprehensive study to built an exact spherically symmetric BH solutions for an emergent gravity framework called symmergent gravity. The gravity theory emerges from quantum loops 
of the underlying QFT, where both the Newton constant $G_N$ and the quadratic curvature coefficient $c_0$ are loop-induced quantities. The symmergent BH differs from the Schwarzschild solution by loop factor $c_0$, which is proportional to the number difference between the fermions and bosons in the theory.  The symmergent BH reduces to Schwarzschild solution, AdS/dS solution or Schwarzschild-AdS/dS solution for different parametric limits of the loop factor constant $c_O$.

To determine sensitivity of the BH properties on the underlying QFT (or emergent nature of gravity), we have studied various features of the symmergent BH solution. We have formed 1 table and 12 figures to display our findings.
From Fig. \ref{fig:brvsr} one can see that the underlying QFT directly affects the horizon radius. Also, as in the Schwarzschild case, the Kretschman scalar diverges at the origin due to not the symmergent gravity contribution but the Schwarzschild piece. From the numerical plot in Fig. \ref{fig:temp} it follows that this the Hawking temperature is a sensitive probe of $c_O$.  Similar size of $c_O$ sensitivity is observed also in the BH entropy (Fig. \ref{fig:entropy2}),  shadow angular radius (Fig. \ref{fig:rovst}), and photon deflection angle (Fig. \ref{fig:defangle}).  These  quantities are sensitive probes of the number difference between the bosons and fermions in the underlying QFT. The QFT here is presumably the standard model plus physics beyond the standard model wherein which dark matter, dark energy and possibly more are contained.  

In section V, we have studied the shadow of the symmergent black hole by using the null geodesics method. We show how the loop factor parameter $c_O$ affects the angular shadow radius $\Theta$ for a static observer at different locations $r_0$ in Fig. \ref{fig:rovst}. The BH shadow angular radius get bigger for increasing the negative value of $c_O$, conversely, BH shadow angular radius shrinks for the positive value of loop factor $c_0$, hence it is concluded that at a larger static observer distance $r_0$, shadow angular radius
$\Theta$ depends on the value of the loop factor $c_0$.

Next, we have studied the gravitational lensing by symmergent BH using Ringdler-Ishak method to check the effect of the loop factor parameter $c_O$ on the weak deflection angle. The weak deflection angle of photon $\hat{\alpha}$ for different loop factor parameter $c_O$ values is plotted in Fig. \ref{fig:defangle}. It is clear that the positive/ negative parameter $c_O$ decreases/ increases with the weak deflection angle $\hat{\alpha}$ as seen in Fig. \ref{fig:defangle}. We show the deviation from general relativity: when loop factor parameter $c_O>0$, the weak deflection angle $\hat{\alpha}$ is smaller than it by Schwarzschild black hole; when $c_O<0$, the weak deflection angle $\hat{\alpha}$ is larger than it by Schwarzschild black hole. The results show that for the increasing impact parameter $R$, weak deflection $\hat{\alpha}$ increases for Schwarzschild-AdS BH, and decreases for the Schwarzschild-dS BHs. The dependence of the deflection angle on $c_O$ in this figure reveals that deflection direction gets reversed with the reversal of the $c_O$ sign at large $R$. Hence, this sensitivity to the sign of $c_O$ shows that the bending of light around the BH is a sensitive probe of the number difference between the bosons and fermions in the underlying QFT.  

The question of whether the BH shadow angular radius changes in the presence of a medium around a gravitating body is a relevant one and we performed this analysis in the last part. Our results show that under plasma medium as $k$ decreases, $\Theta$ decreases appreciably for symmergent gravity. It turns out that plasma have a impact on the BH shadow in symmergent gravity as seen in the effect of the plasma (various $k$ values) depicted in Fig. \ref{fig:k1} for symmergent gravity with $k=1$, in Fig. \ref{fig:k2} for symmergent gravity with $k=0.6$, and  in Fig. \ref{fig:k3} for symmergent gravity with $k=0.2$.  

 In this paper what we get is an AdS/dS type black hole solution in symmergent gravity as $f(R)$ gravity with the effective cosmological constant depending on the model parameters such as the loop factor $c_O$. The effective cosmological constant is something that can be tested in experiments. Moreover, de Sitter spacetime supports the cosmic inflation and  de Sitter universe is a well-structured model of the late universe. The dS type black hole solution can be thought of as the primordial black hole or black hole in the late stages of the universe. It can be detected in near future experiments. Here, by using this model, we show that cosmological constant drives the cosmological expansion, and black hole shadow carries the imprints of the cosmological expansion.

Recently, EHT has been able to measure polarisation, a signature of magnetic fields of the edge of a black hole
\cite{1854874,Akiyama:2021tfw}, it is expected that in near future EHT would able to detect the imprints of dark matter on black hole shadow and in long term, it would achieve the resolution required to observe the influence of emergent gravity on black hole shadow. Further research can thus shed light on the modified gravity theories such as emergent gravity.

With regards to future work, it would be interesting to construct a stationary black hole solution and study its shadow and quasinormal modes that may shed some light on possible signature of the existence of the emergent gravity in the observations of EHT or LIGO. Besides, measurement of the model parameters can reveal the structure of the matter sector.  For instance, a measurement of $c_O$  gives the number difference between the bosons and fermions in the matter sector (a renormalizable QFT comprising the known matter). This field-theoretic meaning is not something does not exist in other eemergent gravity theories. Let us suppose that future data (from the EHT or others) prefer a specific value $c_O^{(exp)}$. Then, symmergent gravity immediately fixes boson-fermion number difference to be $(n_B-n_F)^{(exp)}= 2048 c_O^{(exp)}$.  This experimental value will imply new physics beyond the Standard Model, and the implied new particles can be searched at collider experiments like the LHC experiments. This gravity-matter concord is not present in other gravity theories, and it is in this sense that symmergent gravity is distinguished to have physical implications for both the gravity and matter sectors \cite{demir1,demir2,demir3}. 

\acknowledgements
The authors are grateful to the Editor and anonymous Referee for their valuable comments and suggestions to improve the paper.

\end{document}